# Effect of Isopropanol on Gold Assisted Chemical Etching of Silicon Microstructures

L. Romano[a,b,c], J. Vila-Comamala[b,c], K. Jefimovs[b,c], M. Stampanoni[b,c]

[a] Department of Physics and CNR-IMM- University of Catania, 64 via S. Sofia, Catania, Italy;
[b] Paul Scherrer Institut, 5232 Villigen PSI, Switzerland;
[c] Institute for Biomedical Engineering, University and ETH Zürich, 8092 Zürich, Switzerland

**Abstract**

Wet etching is an essential and complex step in semiconductor device processing. Metal-Assisted Chemical Etching (MacEtch) is fundamentally a wet but anisotropic etching method. In the MacEtch technique, there are still a number of unresolved challenges preventing the optimal fabrication of high-aspect-ratio semiconductor micro- and nanostructures, such as undesired etching, uncontrolled catalyst movement, non-uniformity and micro-porosity in the metal-free areas. Here, an optimized MacEtch process using with a nanostructured Au catalyst is proposed for fabrication of Si high aspect ratio microstructures. The addition of isopropanol as surfactant in the HF-$H_2O_2$ water solution improves the uniformity and the control of the $H_2$ gas release. An additional KOH etching removes eventually the unwanted nanowires left by the MacEtch through the nanoporous catalyst film. We demonstrate the benefits of the isopropanol addition for reducing the etching rate and the nanoporosity of etched structures with a monothonical decrease as a function of the isopropanol concentration.

**Keywords:** silicon microfabrication, Au nanostructures, X-ray gratings, anisotropic wet etching, metal assisted chemical etching

Corresponding author: Lucia Romano, lucia.romano@psi.ch

1. **Introduction**

Fabrication of high-aspect-ratio silicon micro- and nanostructures is a key process in many applications, such as microelectronics [1], microelectromechanical systems [2,3], sensors [4], photonic devices [5], thermoelectric materials [6], battery anodes [7], solar cells [8], and X-ray optics [9]. Microfabrication is usually achieved by reactive ion etching [10], which requires high investment in tools and maintenance. Anisotropic wet chemical etching methods, such as the KOH-based etching [11,12], have been used for fabrication of grooves and pores in Si at micro- and nanoscale. However, the aspect ratio of etched trenches is limited by the etching rate ratio between different crystallographic orientations and only possible in simple geometries like linear gratings or crossed linear gratings defined by the direction of <111> crystallographic planes of Si.

As an alternative approach for fabricating Si microstructures, Metal Assisted Chemical Etching [13] has attracted great interest [14] because of its simplicity, low fabrication costs, and ability to generate high aspect ratio nanostructures such as nanowires and nanoholes [15]. Several acronyms were reported for this process, since 2015 the community seemed to agree with the common acronym of "MacEtch", which will be used here. Unlike KOH wet etching [12], the MacEtch process is independent of crystal orientation and may be used to create a wide variety of hole profiles, trenches, morphologies and paths. An advantage of the method is the considerable reduction in fabrication costs and complexity. MacEtch has been successfully applied for X-ray optics fabrication for nanoscale patterns for synchrotron-based X-ray imaging methods [16,17]. For X-ray grating interferometry imaging, the fabrication of Si microgratings requires sharp vertical profiles, high aspect ratios, high accuracy of pitch size and duty cycle, and finally uniformity over large area. These requirements are especially stringent for X-ray medical diagnostics for which larger field of view are a must. Thus gratings require microfabrication on area of many squared centimetres [9], with aspect ratio and pitch size that depend on the used energy, specific design and performances (pitch size in the range of 1-20 µm, aspect ratio in the range of 10 - 100).

In MacEtch, a metal layer (e.g. gold, Au) is patterned onto the substrate (e.g. Si) to locally increase the dissolution rate of the substrate material in an etchant solution including a fluoride etchant such as hydrofluoric acid (HF) and an oxidizing agent such as hydrogen peroxide ($H_2O_2$). Au is used as preferential catalyst for Si high aspect ratio structures because it is chemically stable, it does not oxidize and it has one of the fastest etching rates [18]. $H_2O_2$ is reduced at the Au surface producing water and at the same time injecting holes through the Au catalyst into Si. At the Au–Si interface, the Si atoms underneath the Au catalyst are oxidized by the holes and dissolved in the HF solution as $H_2SiF_6$. $H^+$ ions and hydrogen gas ($H_2$) are also produced as byproducts of the reaction. Porous metal catalyst film has been reported [2,19,20] to improve the etching performances of micro-scaled Si trenches since the porous film favors the reactant diffusion allowing a uniform etching rate of patterns in the scale of micrometers. Therefore, two regimes can be distinguished in literature [2,14,19]: i) nanoscale patterns, in which the etchant species diffuses through the pattern edges, and ii) microscale patterns with nanoporous films, in which the porosity of the film itself controls the diffusion length. In both regimes, the catalyst geometry significantly affects the etching performance.

Moreover, charge carriers are injected into Si and charge distribution affects the catalyst movement [21], so that parallel and elongated structures [16] are more difficult to etch than spaced cavities [22]. Electron-hole concentration balancing structures were used to achieve a vertical etch profile [16], negative carbon mask [23,24], electrical bias [24,25], magnetic catalyst [26] to improve the control of the catalyst movement [27-29]. The holes diffuse from the Si under the metal to either the off-metal areas or to the walls of the pore if the rate of the holes consumption at the Au–Si interface is smaller than the rate of holes injection. Accordingly, the off-metal areas or sidewalls of the pore may be etched and form microporous silicon, analogous to the case of electrochemical or stain etching [15].

Many studies have been done on surface morphologies and etch rates of Si surface in TMAH or KOH aqueous solutions with addition of alcohol (see for example [30] and reference therein). Though the alcohol does not take directly part in the etching process, it strongly affects the etching results. Both etch rate and roughness of etched surface depend on the alcohol concentration in the etching solution, which is connected with the adsorption phenomena on the etched surface. Isopropanol (IPA) [20] and ethanol [5] have been used as surfactant in MacEtch solutions. While systematic observations about KOH and TMAH etching as a function of IPA concentration and models based on surface tension measurements were reported, a similar study is missing for MacEtch process. In this paper, we characterized the etching rate and the microporosity of the Si trenches produced by MacEtch as a function of IPA concentration in the $HF-H_2O_2$ solutions. The sidewalls of the Si trenches are contoured by a nanoporous layer, whose thickness depends on the ratio between IPA and $H2O2$ in the etching solution.

We used the previously reported self-assembly nanostructuring of the metal film [20] to improve the metal pattern stability during the etching in order to realize high-quality vertical Si microstructures in the microscale regime. Si grating microstructures with high aspect ratio and pitch size in the range of 2.4 – 23 μm have been successfully realized with very smooth surface so to be used as X-ray optics components [16]. A full protocol for fabrication of Si micro-gratings is proposed in section 3.1. The use of IPA additive in MacEtch solution improved the uniformity of the etching over large area and reduced the undesired off-metal area porosity. Finally, the characterization of the IPA effect is reported in section 3.2.

2. **Material and methods**

P-type <100> 4 inches Si wafers with resistivity of 1-30 $\Omega$cm were used in this work. The details of conventional UV photolithographic process with positive photoresist, the Au thin film deposition (10$\pm$1 nm) and lift-off, are reported elsewhere [20]. We fabricated several Au grating patterns with pitch size of p = 2.4, 4.8, 6 and 23 μm and duty cycle of 0.5. We realized the Au nanostructuring self-assembly by annealing in the temperature range 180-250 °C for 30 min in air - using a standard hot-plate. MacEtch was performed by dipping the Au patterned Si into the etching solution, the details are reported elsewhere [20]. Some samples were etched as full wafers (grating pattern was 70x70 mm$^2$); some others were cleaved in 20x20 mm$^2$. No relevant differences were observed in the etching rate or morphology as a function of the etched area. HF (10 vol %), $H_2O_2$ (30 vol%), IPA (VLSI grade) and KOH were purchased from Microchemicals GmbH. Residual nanowires were eventually etched away after MacEtch by dipping the sample into KOH bath (KOH at 2 w% in deionized water at room temperature) and then rinsing in deionized water.

The Si etching rate depends on the relative ratio between the concentration of the etchants HF and $H_2O_2$. We used the compact notation by Hildreth et al. [27] to indicate the experimental conditions of the etching with the following details: $x=\rho^y$, where $\rho=[HF]/([HF]+[H_2O_2])$ and y= [HF], [HF] and $[H_2O_2]$ are the concentrations of HF and $H_2O_2$, respectively, in moles per liter.

Scanning Electron Microscopy (SEM) was performed with a Zeiss Supra VP55 – Gemini column with In-lens Everhart-Thornley detector and 30 µm aperture. SEM cross sections were obtained by cleaving the Si substrates and positioning the cleaved surface at 85 deg with respect to the SEM beam in order to have a partial view of the grating surface. High resolution (HR) images were realized by using 7.5 µm aperture for the best resolution, 10 kV acceleration voltage and working distance of 3.5 mm to maximize the signal-to-noise ratio. In HR images the signal has been optimized by positioning the sample with the cleaved surface at 90 deg with respect to the SEM beam and titling the sample along the cleaved surface (<2.5 deg) in order to have the same edge enhancement effect on both sides of the Si trenches. This helps to minimize the noise in the contrast profile of the edge of Si trench.

3. **Results & Discussion**

3.1 **Micrograting fabrication**

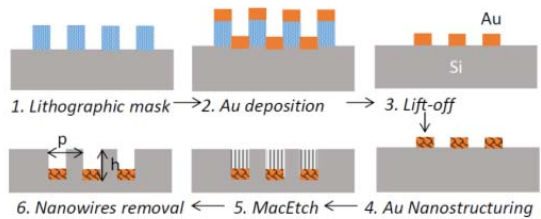

**Figure 1.** Schematic (not to scale) of the full fabrication process: 1) sacrificial photolithographic mask; 2) thin Au deposition; 3) lift-off; 4) Au nanostructuring by annealing; 5) MacEtch; 6) etching of residual nanowires. In color image: photoresist is in light blue, Au is in orange, Si is in grey.

The full fabrication process is schematically reported in Figure 1: 1) a sacrificial photoresist pattern was formed on Si substrate by conventional photolithographic method; 2) the substrate is cleaned with Oxygen plasma to form an Oxygen-terminated Si surface and a thin film of Au is subsequently deposited on Si; 3) lift-off in the proper solvent leaves the Au pattern on the substrate and removes the photoresist; 4) Au nanostructuring by annealing (30 min in air); 5) MacEtch by dipping the substrate in an etching solution containing HF and $H_2O_2$, resulting into the Au pattern that sinking into the Si substrate; 6) residual nanowires removal in KOH or HF solution.

Gold nanostructuring self-assembly was realized by thermal de-wetting of the metal on Oxygen-terminated Si surface [20]. In contrast to thin film deposition, where the Au porosity depends on various deposition parameters (substrate temperature, deposition rate, film thickness and deposition method), the porosity of de-wetted Au can be finely tuned as a function of the annealing temperature and the film thickness. An example of metal de-wetting is reported in Figure 2. The Au coverage as a percentage of the full area of the sample is reported as a function of the annealing temperature for isochronal treatment of 30 min. For each de-wetting temperature, the area was measured by detecting the contrast variation (Au corresponds to the bright signal) in several SEM images at a magnification of 200 kX. Some examples of SEM images in plan view of the metal pattern at different temperatures are shown in insets of Figure 2. The metal coated area decreases as the annealing temperature increases. Once de-wetted, the metal pattern is more resistant to mechanical instabilities and acts like a membrane allowing the etching solution to reach the silicon surface [20].

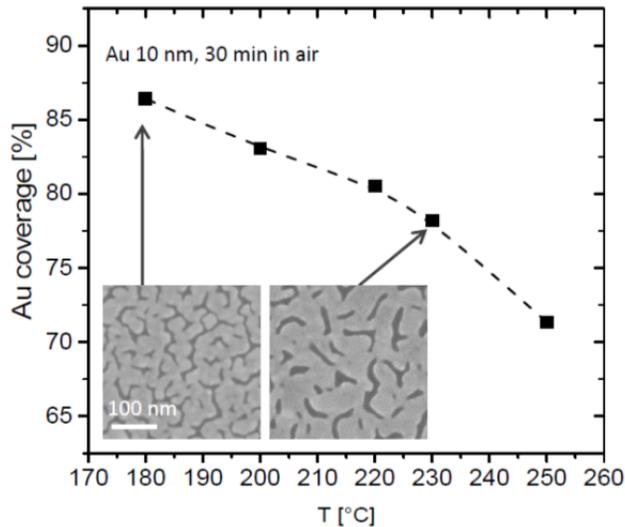

**Figure 2.** Au coverage in % of the surface area measured in SEM images in plan view as a function of the annealing temperature. The Au film thickness was 10 nm and the annealing was performed in air for 30 min. Insets show SEM images of Au film annealed at 180°C (left) and 230°C (right), the scale marker is the same in both images.

MacEtch process is a fragile equilibrium of chemical reactions and mass transport. The porous catalyst layer improves the diffusion of the reactants through the catalyst [19] keeping the etching rate uniform in the catalyst area and preventing the edges bending of the Au strips due to a faster etching at the borders [19]. However, if the Au coverage is too small, the catalyst film could form a disconnected network, where pieces of catalyst can be detached during the etching reaction and be subjected to out-of-vertical path, as reported by Hildreth et al.[2]. Therefore the annealing process has to be tuned to maintain the metal film a continuous network and the desired resolution at the line edges of the patterned catalyst.

The concentrations of $H_2O_2$ and HF affect not only the etching rate, but also the morphologies of the etched structures. As the composition varies from high to low $\rho$, mesopores, cone-shaped macropores, craters and eventually smooth surfaces are obtained [18]. In a previous work [20], we studied the etching rate for a fixed Au film thickness of 10 nm, the etching rate monotonically decreases as $\rho$ increases (an example of the measured etching rate in presence of HF and H2O2 is reported in the inset of Figure 4.a). For a fixed $\rho$, the etching rate speeds up with the increase of [HF]. With high $\rho$ the etching is more vertical, for example nanowires are usually realized with 0.9<$\rho$<1 [15]. For microstructures, we determined [20] a good $\rho$ with an intermediate value ($\rho$~0.6) that guarantees a sustainable etching rate with a good verticality. Some examples of etched gratings are reported in Figure 3.a and 3.b.

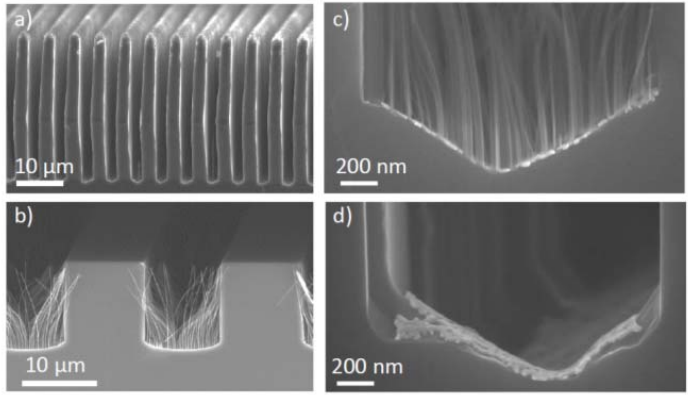

**Figure 3.** SEM in cross section of several gratings with different pitch size p: a) 6 μm (190°C, x=0.60$^{3.50}$) and HF dip of 20 min to remove nanowires; b) 23 μm (250°C, x=0.92$^{4.98}$); c) and d) 4.8 μm (180°C, x=0.60$^{3.50}$). Nanowires are formed in b) and c) due to the Au mesh pattern. Nanowires were removed after the MacEtch by dipping the sample in HF 10% for 20 min (a) or KOH solution (2 w% in water) for 2 min (d). The effect of KOH on the Si trenches is visible in the different angle of the trench base in d) with respect to c).

When high ρ MacEtch solutions are used in combination with high temperature of Au de-wetting, the Au membrane can leave nanowires at the bottom of the grating, as it can be observed in Figure 3.b (annealing T=250°C, x=(0.92)$^{4.98}$). These nanowires can be detrimental for the following filling of the Si trenches with high absorption x-ray material, such as Au electroplating [31,32] or casting [33]. For ρ<0.6 the nanowires can be also consumed by the MacEtch solution itself, being very thin Si structures that are easily oxidized by $H_2O_2$ and etched away by HF. Nanowires are usually observed only very close to the bottom of the grating, as it is showed in Figure 3.c. These residual nanowires were etched away after MacEtch by dipping the sample in HF water solution (10%). Air exposure for few hours after MacEtch oxidizes the nanowires and a complete removal can be obtained in 20 – 30 min, depending on the oxidation of the nanowires and their size. An example of a Si grating in which the nanowires have been removed by HF is shown in Figure 3.a. Otherwise, nanowires were fast etched away by dipping the sample into a high-diluted KOH solution (2 w% in deionized water) at room temperature. For example, 2 min the etching can efficiently remove the nanowires with a low damage to the Si trenches. The effect of the KOH process for the complete removal of the nanowires can be compared in the images before (Figure 3.c) and after the process (Figure 3.d).

### 3.2 IPA effect on MacEtch process

Another issue of MacEtch is the $H_2$ gas release during the etching process. The $H_2$ is produced as a by-product of reaction [15] and it can substantially affect the etching results since very large bubbles can be formed on the surface of the grating, dramatically preventing a uniform etching. We observed with the naked eye several bubbles sticking on the grating surface just after few minutes etching. Later, at the end of the MacEtch process, we detected several areas of comparable size with a different contrast in optical microscope. This phenomenon appeared to be much more critical in patterned microstructures than mesh pattern for nanowires since the gas bubbles can be stabilized in the etched structure with liquid solution exhibiting the Cassie–Baxter wetting state [34,35]. The resulting etching dis-uniformity is shown in Figure 4.a, where a region of at least 50 μm in diameter is etched for a smaller depth compared to the surrounding area. In order to prevent the effect of the $H_2$ gas, we added the IPA to the etching solution. IPA is completely miscible with HF and lowers the surface tension of the liquid solution, favoring the formation of smaller gas bubbles. It has been reported for KOH etching that the surfactant forms a layer physically covering the surface and prevents the formation of large $H_2$ bubbles, reducing the amount and size of etchant inhomogeneity in contact with the surface. As a result, the surface roughness is reduced and the morphology is smoothed [36]. With the addition of IPA, we observed the gas release in very small bubbles, almost invisible with the naked eye. Therefore, it is much more difficult for the gas bubbles to stabilize on the grating surface during the etching. We obtained a better uniformity of the etching depth on a large area, up to 70x70 $mm^2$ area [33]. An example is reported in Figure 4.b, where a grating is etched with the same conditions of Figure 4.a with the addition of IPA in the MacEtch solution.

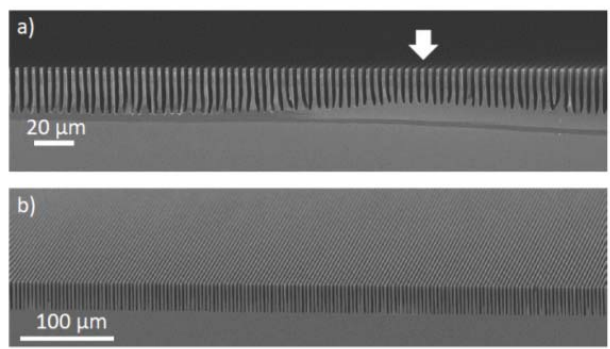

**Figure 4.** SEM in cross section of 4.8 um pitch grating etched with regular MacEtch a) and MacEtch solution with the addition of IPA. The arrows indicate the presence of a gas bubble preventing the uniform etching of the grating.

We observed that the addition of IPA substantially affects the etching rate, in the following we report a detailed study of the IPA effect on the etching rate in relationship to the other parameters of the MacEtch process.

For a given catalyst material and morphology, the MacEtch etching rate is usually characterized as a function of $\rho$ [18]. Since for water diluted HF at 10 vol%, the molar concentration is [HF]~5 mol/l. We added the deionized water to HF and $H_2O_2$ in the MacEtch solution to vary $\rho$ in the range 0-1. We substituted the quantity of deionized water with IPA in order to quantify the effect of IPA on the etching rate as a function of $\rho$. The etching rate vs $\rho^{2.4}$ in pure HF-$H_2O_2$-water solutions (black squares) or with IPA (red dots) is reported in Figure 5.a. The etching rate is strongly reduced by IPA with a smaller dependence on $\rho$ with respect to the HF-$H_2O_2$-water solution. Therefore, within the range of $\rho$ between 0.4 and 0.8, we can study the dependence of the etching rate on the IPA concentration only. We prepared the etching solution by adding the IPA volume to the HF-$H_2O_2$-water solution and we calculated the relative IPA concentration in mol/l. Figure 5.b reports the etching rate as a function of the IPA concentration in the etching solution, $\rho$ is in the range 0.4-0.8 and is shown as label in Figure 5.b.

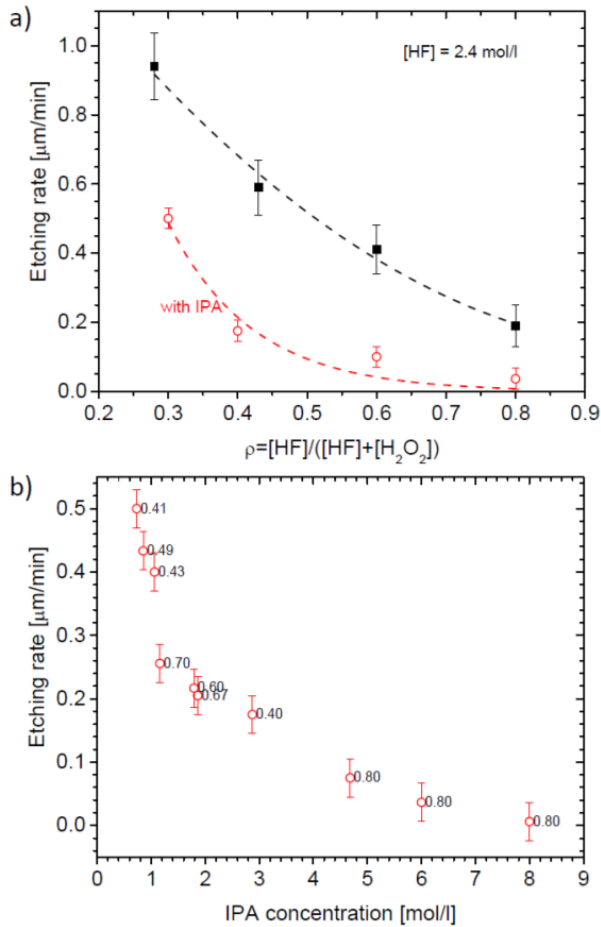

**Figure 5.** MacEtch etching rate as a function of $\rho$ (a) for HF-$H_2O_2$-water (back squares) and HF-$H_2O_2$-IPA (red dots) solutions, and IPA concentration (b) for $0.4<\rho<0.8$, labels in b) indicates the $\rho$.

This kind of plot can be directly compared to the KOH etching rate as a function of IPA concentration with a fixed KOH concentration. In KOH etching, the etching rate is controlled by the KOH concentration, while in MacEtch the same role is played by $\rho$. Within the experimental errors, Figure 5.b indicates that for $0.4<\rho<0.8$ the etching rate is mainly controlled by the IPA concentration. Moreover, in KOH etching the etching rate has a minimum as a function of the IPA concentration [37], while in MacEtch the etching rate seems to decrease until the etching is eventually completely stopped.

There is another phenomenon that is very sensitive to the IPA concentration. Microporosity is due to the diffusion in off-metal areas of the injected holes in the semiconductor. Lianto et al. [38] reported the formation of micro-pits in MacEtch with compact Au catalyst and the density of micro-pits increased with increasing the $H_2O_2$ content in the solution. We observed the formation of micro-pits (cavities with radius > 100 nm) on the top of the etched gratings and up to an average depth of 2-3 µm from the top surface. In addition to micro-pits [38], nanoporosity (pore size ~ 10 nm) has been reported [3,39]. By HR-SEM we detected the presence of a continuous layer of nanoporous Si, with thickness ranging between 250 - 150 nm for $\rho$= 0.37 and $\rho$=0.8, respectively. The nanoporous layer can degrade the grating surface for the following treatments that are required to finally realize X-ray absorption gratings, such as Au electrodeposition or metal conformal coating [33]. We observed that the nanoporous layer contouring the Si etched structures was substantially reduced in presence of IPA in the etching solution. Figures 6.a-c report the top of etched gratings in high magnification SEM. In figure 6.a the grating was etched with $x=0.8^3$ for 40 min (h~15 µm). The continuous nanoporous layer has at least a thickness of 144 nm. Some micro-pits are also visible and they are probably formed as a consequence of agglomeration of smaller pores. Figure 6.b shows similar grating of Figure 6.a etched with a similar solution ($x=0.8^3$, 40 min, h~3 µm) containing IPA with a molar concentration of [IPA]=1 mol/l. With an improved sharpness of the etched profile (less and smaller micro-pits), the addition of IPA in the MacEtch solution reduced the nanoporous layer thickness to 20 nm. Even if the etching velocity decreases of a factor 5, the thickness of the nanoporous layer is reduced by a factor of 7.5. In order to measure the porous layer thickness with a good accuracy, we sampled the layer thickness by using HR-SEM and measuring the line profile in several points of the grating. An example of HR-SEM with magnification of 800 kX is reported in Figure 6.c. The line profile is reported in Figure 6.d and indicates that the signal-to-noise ratio is enough to measure the thickness with enough resolution, even some single nanopores can distinguished. The thickness was measured as FWHM of the contrast profile, the standard deviation over 30 measured profiles in HR-SEM is 1.4 nm with a maximum difference of 2.5 nm between the smallest and greatest measured profile. The results of serval HR-SEM images are reported in Figure 6.e and error bars are ±2.5 nm. Since the Si porosity is related to the injected holes from the $H_2O_2$ reaction with the catalyst, Figure 6.e reports the nanoporous layer thickness as a function of the ratio of IPA and $H_2O_2$ molar concentration to take into account both parameters. A monothonical decrease of the nanoporous layer thickness is observed, indicating once more the passivation effect of IPA between the Si etched surface and the etching solution. As for the etching rate, no local minimum is observed, confirming that the role of IPA in MacEtch is slightly different in comparison to the other wet etching processes [30,37]. The nanoporous thickness was observed to be stable, maintaining the reported values of Figure 6.e even after several hours of etching.

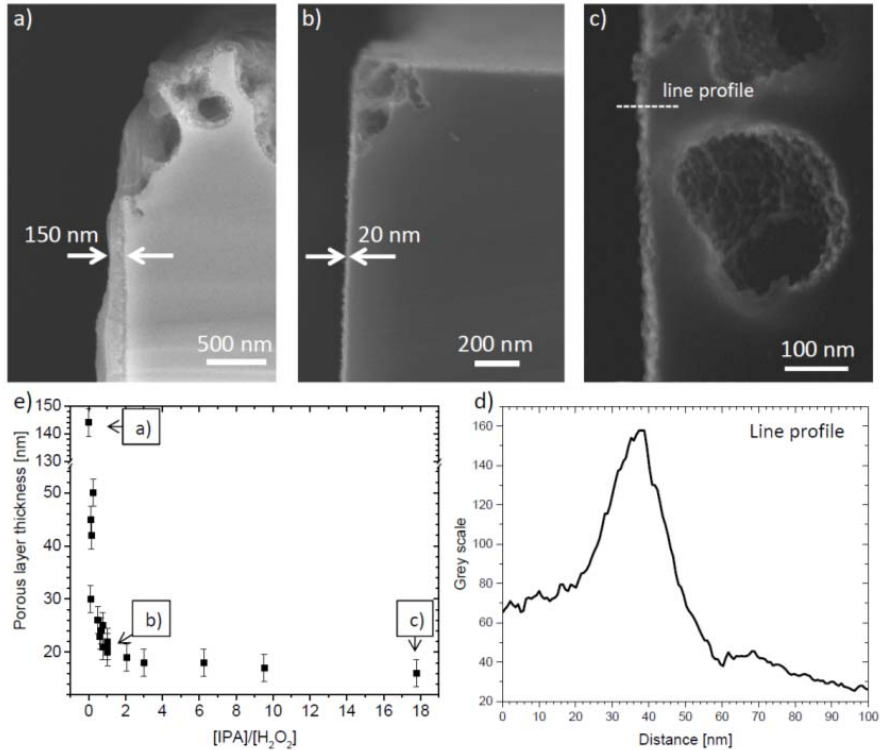

**Figure 6.** SEM in high magnification of the top of the grating with pitch size p = 4.8 μm etched in MacEtch solution with x=0.8[3] for 40 min in presence of deionized water (a) or IPA (b). Micro-pits are present on the top of the grating in both images with size > 100 nm, the nanoporous layer (pore size ~10 nm) was characterized by using HR-SEM (c) with contrast line profile (d). The nanoporous layer thickness (the SEM images of gratings are indicated) as a function of the ratio of molar concentration of IPA and $H_2O_2$ in the etching solution.

According to the literature, in a wet-etch process in presence of surfactant, the surfactant is adsorbed at the silicon–etchant interface, acting as a filter that moderates the surface reactivity by reducing the number of reactant molecules that reach the surface and affecting the number of reactions that are carried out [36]. For IPA in KOH etching the variation of the etching rate with IPA concentration has been explained as following: 1) when IPA concentration is relatively low, the alcohol molecules are adsorbed with their hydrophobic parts on the silicon–solution interface; 2) when the alcohol concentration increases, the (100) surface is attacked by micellar aggregate formed in the solution, while the (110) adsorbs the alcohol molecules from the solution; 3) the (100) surface is exposed to the reactive molecules from the etching solution and its etch rate increases, while the alcohol molecules remain adsorbed on the (110) surface, reducing its etch rate [37]. In MacEtch, the mechanism is more complex, two main things have to be considered to model the MacEtch process of high aspect ratio structures: 1) once the Si trench is created there is a clean Si surface available for further reactions; 2) there is a metal catalyst so the etching rate is measured as a function of the catalyst deepening into the Si substrate.

In HF solution, Si dissolution can occur through direct dissolution reactions:

$$Si + 4h^+ + 4HF \rightarrow SiF_4 + 4H^+ \quad (1)$$

$$SiF_4 + 2HF \rightarrow H_2SiF_6 \quad (2)$$

$$Si + 4HF^{-2} \rightarrow SiF_6^{2-} + 2HF + H_2 \uparrow + 2e^- \quad (3)$$

or via oxidation of Si and subsequent dissolution of $SiO_2$ in HF:

$$Si + 2H_2O \rightarrow SiO_2 + 4H^+ + 4e^- \quad (4)$$

$$SiO_2 + 6HF \rightarrow H_2SiF_6 + 2H_2O \quad (5)$$

It has been reported [40] that the addition of surfactant in buffered HF etching reduces the Si dissolution because the oxidation process of Si (reactions 4 and 5) is suppressed by the adsorption of the surfactant molecules on the Si surface. Moreover, it is well known that IPA reduces the etching of $SiO_2$ in HF [41] because the HF dissociation decreases. The addition of $H_2O_2$ accelerates the HF dissociation and $HF^{2-}$ formation reactions [41], and thus increases the etch rate of oxide. Therefore, the reduction of nanoporous layer contouring the Si trenches can be due to the reactions 4 and 5. The micropits do not seem to be strongly affected by the IPA, as shown in Figure 6.b and 6.c., they are still present in IPA solutions with size in the range of hundreds of nm. The contours of micropits are also nanoporous and they become more defined in presence of IPA because the nanoporous layer thickness is reduced by the IPA. According to our observations, we can infer that the nanoporous layer is related to the reactions 4 and 5 and therefore it can be affected by the IPA, while the nature of micropits could be related to the reactions 1-3.

The alcohol concentration in our experiment is much higher than reported for KOH etching [37], in which the etch rate reduction is observed in [IPA] range of 0.1-2 mol/l. In our experiment the surfactant can be adsorbed on a bigger Si surface with respect to that in Si KOH experiment. Since the formation of microtrenches, for an aspect ratio of 10:1 the Si available surface can be one order of magnitude bigger than a flat surface.

We used a Au nanporous film as MacEtch catalyst, Au is a very common catalyst for organic reactions [42]. Au is reported to be very active into break C-H bonds but catalytic oxidation of IPA with Au -based catalysts has been reported for temperature in the range of 100 °C [43]. Therefore, at room temperature and in dark conditions as in our experiment [44], a catalytic oxidation of IPA should be excluded. Additional results with other catalyst are needed to properly evaluate the chemical catalyst-IPA interaction. IPA can be adsorbed on the metal surface and affect the interaction of the catalyst with the etching solution with the reduction of the etching rate. This effect can be present and contribute to the slowing down of the etching process, together with the IPA reduction of $SiO_2$ etching as anode reaction [45].

4. Conclusions

We demonstrated the fabrication route of Si micro-gratings with pitch size in the range of 2-23 μm by using a combined approach of nanostructured metal pattern (thermal de-wetting of Au thin film) and MacEtch etching. Gas bubble release that hinders the etching uniformity over large areas, and the undesired microporosity in the off-metal areas were significantly reduced by adding IPA to the etching solution. The etching rate decreases as a function of the IPA concentration as for KOH etching but apparently without minimum. Moreover, the nanoporous layer thickness reduces from 150 nm to 20 nm as a function of the ratio of IPA and H2O2 concentrations. This kind of porosity seems to be related to the Si dissolution via the oxidation reaction and subsequent HF dissolution of oxide, while the micro-pits can be related to the direct Si dissolution reactions. No constrains of the reported methodology appeared about the etching of grating with large pitch size (2.4-23 µm) and depth (80 µm), opening new possibilities of fabricating high aspect ratio Si microstructures.

**Acknowledgements**

This work has been partially funded by the ERC-2012-STG 310005-PhaseX grant, ERC-PoC-2016 727246-MAGIC grant and the Fondazione Araldi Guinetti. We would like to thank C. David and V. Guzenko from PSI-LMN for collaboration and valuable discussion.